\documentclass[twoside,twocolumn,9pt]{article}
\usepackage{extsizes}
\usepackage[super,sort&compress,comma]{natbib} 
\usepackage[version=3]{mhchem}
\usepackage[left=1.5cm, right=1.5cm, top=1.785cm, bottom=2.0cm]{geometry}
\usepackage{balance}
\usepackage{mathptmx}
\usepackage{sectsty}
\usepackage{graphicx} 
\usepackage{lastpage}
\usepackage[format=plain,justification=justified,singlelinecheck=false,font={stretch=1.125,small,sf},labelfont=bf,labelsep=space]{caption}
\usepackage{float}
\usepackage{fancyhdr}
\usepackage{fnpos}
\usepackage[english]{babel}
\addto{\captionsenglish}{%
  
}
\usepackage{array}
\usepackage{droidsans}
\usepackage{charter}
\usepackage[T1]{fontenc}
\usepackage[usenames,dvipsnames]{xcolor}
\usepackage{setspace}
\usepackage[compact]{titlesec}
\usepackage{hyperref}

\usepackage{epstopdf}

\definecolor{cream}{RGB}{222,217,201}

\begin{document}

\pagestyle{fancy}
\thispagestyle{plain}
\fancypagestyle{plain}{
\renewcommand{\headrulewidth}{0pt}
}

\makeFNbottom
\makeatletter
\renewcommand\LARGE{\@setfontsize\LARGE{15pt}{17}}
\renewcommand\Large{\@setfontsize\Large{12pt}{14}}
\renewcommand\large{\@setfontsize\large{10pt}{12}}
\renewcommand\footnotesize{\@setfontsize\footnotesize{7pt}{10}}
\makeatother

\renewcommand{\thefootnote}{\fnsymbol{footnote}}
\renewcommand\footnoterule{\vspace*{1pt}%
\color{cream}\hrule width 3.5in height 0.4pt \color{black}\vspace*{5pt}} 
\setcounter{secnumdepth}{5}

\makeatletter 
\renewcommand\@biblabel[1]{#1}            
\renewcommand\@makefntext[1]%
{\noindent\makebox[0pt][r]{\@thefnmark\,}#1}
\makeatother 
\renewcommand{\figurename}{\small{Fig.}~}
\sectionfont{\sffamily\Large}
\subsectionfont{\normalsize}
\subsubsectionfont{\bf}
\setstretch{1.125} 
\setlength{\skip\footins}{0.8cm}
\setlength{\footnotesep}{0.25cm}
\setlength{\jot}{10pt}
\titlespacing*{\section}{0pt}{4pt}{4pt}
\titlespacing*{\subsection}{0pt}{15pt}{1pt}

\fancyfoot{}
\fancyfoot[LO,RE]{\vspace{-7.1pt}\includegraphics[height=9pt]{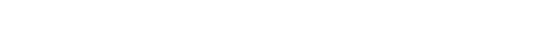}}
\fancyfoot[CO]{\vspace{-7.1pt}\hspace{13.2cm}\includegraphics{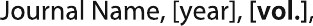}}
\fancyfoot[CE]{\vspace{-7.2pt}\hspace{-14.2cm}\includegraphics{head_foot/RF}}
\fancyfoot[RO]{\footnotesize{\sffamily{1--\pageref{LastPage} ~\textbar  \hspace{2pt}\thepage}}}
\fancyfoot[LE]{\footnotesize{\sffamily{\thepage~\textbar\hspace{3.45cm} 1--\pageref{LastPage}}}}
\fancyhead{}
\renewcommand{\headrulewidth}{0pt} 
\renewcommand{\footrulewidth}{0pt}
\setlength{\arrayrulewidth}{1pt}
\setlength{\columnsep}{6.5mm}
\setlength\bibsep{1pt}

\makeatletter 
\newlength{\figrulesep} 
\setlength{\figrulesep}{0.5\textfloatsep} 

\newcommand{\topfigrule}{\vspace*{-1pt}%
\noindent{\color{cream}\rule[-\figrulesep]{\columnwidth}{1.5pt}} }

\newcommand{\botfigrule}{\vspace*{-2pt}%
\noindent{\color{cream}\rule[\figrulesep]{\columnwidth}{1.5pt}} }

\newcommand{\dblfigrule}{\vspace*{-1pt}%
\noindent{\color{cream}\rule[-\figrulesep]{\textwidth}{1.5pt}} }

\makeatother

\twocolumn[
  \begin{@twocolumnfalse}
{\includegraphics[height=30pt]{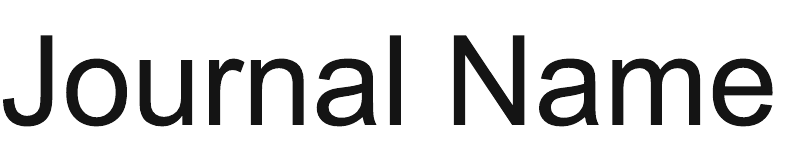}\hfill\raisebox{0pt}[0pt][0pt]{\includegraphics[height=55pt]{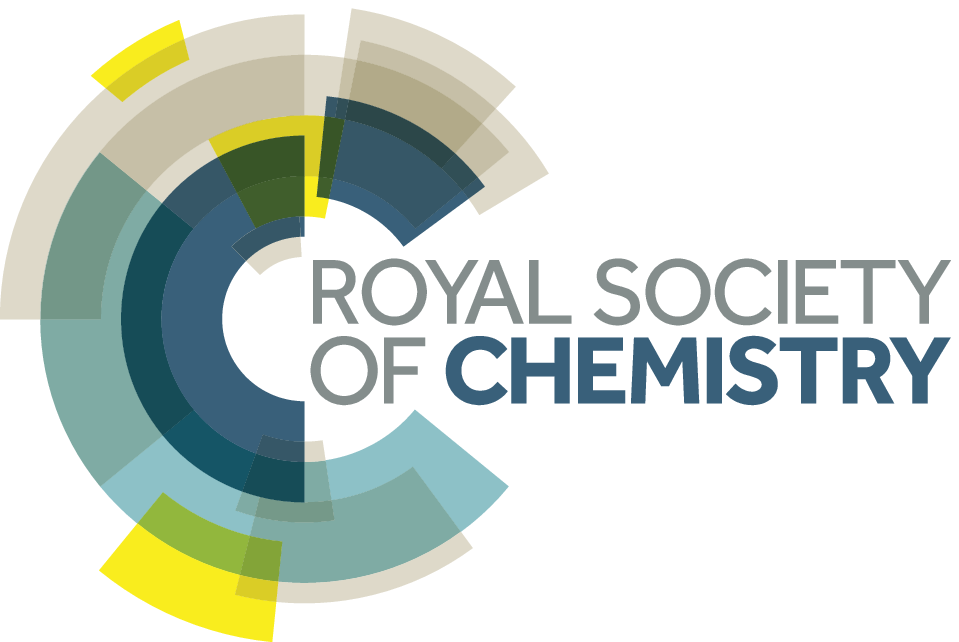}}\\[1ex]
\includegraphics[width=18.5cm]{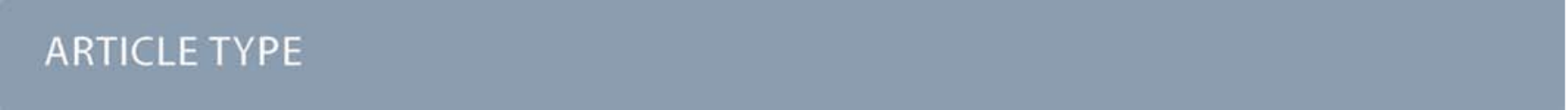}}\par
\vspace{1em}
\sffamily
\begin{tabular}{m{4.5cm} p{13.5cm} }

\includegraphics{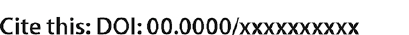} & \noindent\LARGE{\textbf{MXene-based Ti$_2$C/Ta$_2$C lateral heterostructure: an intrinsic room temperature ferromagnetic material with large magnetic anisotropy$^\dag$}} \\
\vspace{0.3cm} & \vspace{0.3cm} \\

 & \noindent\large{S.~\"Ozcan$^{\ast}$\textit{$^{a}$} and B.Biel{$^{b\ddag}$}}\\

\includegraphics{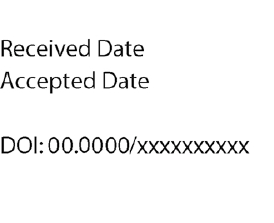} & \noindent\normalsize{Two-dimensional (2D) lateral heterostructures (LH) combining  Ti$_2$C and Ta$_2$C MXenes were investigated by means of first-principles calculations. Our structural and elastic properties calculations show that the lateral Ti$_2$C/Ta$_2$C heterostructure results in a 2D material that is stronger than the original isolated MXenes and other 2D monolayers such as germanene or MoS$_2$. The analysis of the evolution of the charge distribution with the size of the LH shows that, for small systems, it tends to distribute homogeneously between the two monolayers, whereas for larger systems electrons tend to accumulate in a region of $\sim$~6 {\AA} around the interface.
The work function of the heterostructure, one crucial parameter in the design of electronic nanodevices, is found to be lower than that of some conventional 2D LH. Remarkably, all the heterostructures studied show a very high Curie temperature (between 696 K and 1082 K), high magnetic moments 
and high magnetic anisotropy energies. These features make (Ti$_2$C)/(Ta$_2$C) lateral heterostructures very suitable candidates for spintronic, photocatalysis, and data storage applications based upon 2D magnetic materials.} \\

\end{tabular}

 \end{@twocolumnfalse} \vspace{0.6cm}

  ]

\renewcommand*\rmdefault{bch}\normalfont\upshape
\rmfamily
\section*{}
\vspace{-1cm}


\footnotetext{\textit{$^{a}$~Department of Physics, Aksaray University, 68100 Aksaray, Turkey. E-mail: sozkaya@aksaray.edu.tr}}
\footnotetext{\textit{$^{b}$~ Department of Atomic, Molecular and Nuclear Physics \& Instituto Carlos I de F\'isica Te\'orica y Computacional, Faculty of Science, Campus de Fuente Nueva, University of Granada, 18071 Granada, Spain. }}

\footnotetext{\dag~Electronic Supplementary Information (ESI) available: [details of any supplementary information available should be included here]. See DOI: 10.1039/cXCP00000x/}

\footnotetext{\ddag~Additional footnotes to the title and authors can be included \textit{e.g.}\ `Present address:' or `These authors contributed equally to this work' as above using the symbols: \ddag, \textsection, and \P. Please place the appropriate symbol next to the author's name and include a \texttt{\textbackslash footnotetext} entry in the correct place in the list.}



\section{Introduction}

Over the past few years, two-dimensional (2D) materials have experienced a major research effort in materials science. Due to their superb electronic and mechanical properties~\cite{Chen-15, Jin-2016, Mak-10, Jing-13}, many 2D materials such as graphene, black phosphorus, transition metal thin films, hexagonal boron nitride (h-BN), and molybdenum
disulfide (MoS$_2$) have been obtained in experiments and predicted by theory ~\cite{Nicolosi-13, Butler-13, Chen-2017, Pumera-17, Kong-17, Mannix-17, Tan-17}. However, the fulfillment of the applications of 2D materials is often limited by some inherent difficulties, e.g., such as the zero gap in graphene, the low carrier mobility in MoS$_2$, or the instability of phosphorene upon oxidation, among others ~\cite{Mir20}. 

2D materials might show, however, distinct physical and chemical properties when two or more of them are combined into a heterostructure ~\cite{Muhammad21, Zhang22, Bao21, Cui-15}. The formation of heterostructures thus determines in those cases the overall performance of the hybrid material ~\cite{Liu19, Gibertini19, Jin18, Novoselov16}, leading to a new area to study interfaces and device applications ~\cite{Liu18, Pomerantseva17}. Generally, two types of 2D heterostructures can be considered: vertical heterostructures (VH) and lateral heterostructures (LH). In the vertical heterostructures, ~\cite{Geim-13} the 2D monolayers are vertically stacked layer-by-layer, bonded by weak van der Waals (vdW) forces. LH, in contrast, are built by seamless in-plane, side-by-side growth of the 2D monolayers, bonded together by covalent bonds. Since the LH are connected by chemical bonds in the “interline”, they can potentially show enormous epitaxial quality and stability, and exhibit physicochemical properties that might immensely differ from those of the original 2D materials that form the LH ~\cite{ Eda-12, Lin-14, Naylor-17}. Moreover, LH have twice the surface and simpler band alignment compared with the VH ~\cite{Li15, Zhang16}. All of this makes LH with well-controlled domain sizes and atomic sharp interface equally or even more promising than their vertical counterparts ~\cite{Zhao18}. Finding suitable candidate materials to build a LH with the desired properties has some difficulties due to structural limitations, such as the need for a small lattice mismatch of materials in order to make the interface between the two monolayers as seamless as possible. Hence, a theoretical exploration of the LH structural and electronic properties has been essential to overcome these issues prior to the experimental work ~\cite{Sun-16, Fiori-12, Chen-17, Yang-17, Cheng-17, Jin-16}. 

In 2012, Levendorf {\it et al}. \cite{Levendorf12} developed a patterned regrowth method for the synthesis of lateral graphene/h-BN heterostructures on Cu foil. Following this result, several groups \cite{Gong14, Huang14, Duan14} reported the fabrication of LH for transition metal dichalcogenides (TMDs), and since then several experimental reports on the synthesis of various TMD-based LH have been published. Recently, Zeng {\it et al}. ~\cite{Zeng17} prepared
2D lateral WC-graphene heterostructures with excellent chemical stability and reactivity, thus triggering a renewed attention to this field. In particular, unusual electronic and magnetic properties such as the giant magnetocrystalline anisotropy energy (MAE) were demonstrated at the interface as the phosphorene/WSe$_2$ interface ~\cite{Tian16}, which shows the abundant opportunities of LH for applications in spintronics. However, much more work needs still to be done in order to fully exploit the extraordinary potential of magnetic LH.

In this context, 2D MXenes have attracted increasing attention in recent years due to their significant potential applications in electronic devices, sensors, catalysis, energy conversion, and storage systems ~\cite{Shein-12, Kurt-12, Gao-2016, Zha-16, Pandey-17, Fu-17}. After the synthesis of Ti$_3$C$_2$ ~\cite{Naguib-11}, many MXenes have since then been experimentally discovered ~\cite{Hantanasirisakul-18, Naguib-12} or theoretically predicted ~\cite{Wang-17,Anasori-17}. As a typical metallic group of MXenes, M$_2$C(M=Ti, Ta) have also been widely studied, both in experiments and theory~\cite{Naguib-11, Naguib-12, Naguib-13, Wang-14, Yu-17}. The Ti$_2$C and its derivatives, the Ti$_2$CC$_2$, Ti$_2$CO$_2$, and Ti$_2$CS$_2$ MXenes, exhibit good stability and have excellent potential for their use as anode materials due to their strong adsorptions and low diffusion energy barriers, which suggest fast charge-discharge rates ~\cite{Wang-14, Xiao-19, Faraji21}. Besides, although the majority of the functionalized MXene systems are nonmagnetic, the Ti$_{n+1}$X$_n$ (X=C, N) monolayers are magnetic due to the 3d electrons of the surface Ti atoms ~\cite{Sternik18}. On the other hand, Ta$_2$C is stable, has high thermal stability, and provides good electronic conductivity, hence becoming a candidate as an anode material ~\cite{Yu-17}. While the Ta$_2$C is nonmagnetic (NM), its magnetic properties can be tuned by applying mechanical strain ~\cite{Zhao14}. Considering that both materials share the same structural features and present a small lattice mismatch, Ti$_2$C and Ta$_2$C should thus be excellent candidate materials to build magnetic LH. At this point, the critical issue arises of whether LH composed of Ti$_2$C and Ta$_2$C will be stable, and what the actual structural, electronic, and magnetic properties of such LH would be.

Inspired by the aforementioned questions, in this work we provide a comprehensive investigation, based on Density Functional Theory (DFT), into various lateral heterostructures composed by the Ti$_2$C and Ta$_2$C MXene monolayers.

\section{Method}

All the calculations were carried out using the Vienna Ab initio Simulation Package (VASP)~\cite{vasp1,htt, Kre-99, Kres-96}, based on DFT. In this method, the Kohn--Sham single-particle functions are expanded based on plane waves up to a cut-off energy of 450 eV ~\cite{Si15}, found to provide good convergence of the total energy. We have given the convergence results in Supplementary Information (see Table S1.) for spin-polarized calculation. The electron-ion interactions were described by using the projector
augmented-wave (PAW) method~\cite{Kre-99, Bl-94}. For the electron
exchange and correlation terms, the Perdew--Zunger-type
functional~\cite{Per-92,Per-81} was used within the generalized
gradient approximation (GGA)~\cite{Bl-94} in both the PBE parameterization ~\cite{Perdew-1996} and the hybrid exchange–correlation functional (HSE06)~\cite{Heyd-2003, Heyd-2006, Krukau-2006}. Additionally, our calculations were refined using the optimized functional optB86b-vdW ~\cite{Klimes10} to take into account
the van der Waals interactions. Calculations were then carried out
self-consistently using the VASP implementation ~\cite{Klimes11} within the
Roman-Perez and Soler ~\cite{Perez09} scheme. The self-consistent solutions were obtained by employing the
($13\times21\times1$), ($9\times21\times1$), ($7\times21\times1$), and ($4\times21\times1$) Monkhorst--Pack~\cite{Mon-76} grid of k-points for the integration over the Brillouin zone for (Ti$_2$C)p/(Ta$_2$C)q (with p = q = 3, 4, 5, 10, where p and q are the numbers of unit cells of each 2D material repeated to form the LH unit cell), respectively. To prevent spurious interaction between isolated layers, a vacuum layer of at least 15  {\AA} was included along the normal direction to the surface. The elastic tensor of each system is derived from the stress–strain approach ~\cite{Le-02}.

\section{Results and Discussions}

Ti$_2$C and Ta$_2$C have a  trigonal crystal structure (space group P$\overline {3}$m1). The calculated lattice constants of the isolated monolayers Ti$_2$C and Ta$_2$C are 3.05 {\AA} and 3.12 {\AA}, respectively, agreeing well with previous theoretical studies, where a lattice constant of 3.04 {\AA} ~\cite{Shein-12} and 3.14 {\AA} ~\cite{Kurt-12} was found. The lattice mismatch of Ti$_2$C and Ta$_2$C is then merely 2.2 \%, implying that both materials are very suitable to build LH with coherent interfaces. Fig. 1(a) shows the typical structures of our atomic model of the LH. Along the x-direction of the LH in Fig.1(a), single layer Ti$_2$C and Ta$_2$C repeat alternating and periodically, whereas along the y-direction the two different stripes extend to infinity (without any edge). We have named these heterostructures as  (Ti$_2$C)p/(Ta$_2$C)q. To investigate the impact of the LH lateral size we have investigated several LH, for $p$ =$q$ = (2,2), (3,3), (4,4), (5,5), and (10,10). The lateral size of each one of these LH unit cells is then 12.15, 18.25, 24.35, 30.44, and 60.94 {\AA}, respectively.

\begin{figure}[h] \centering
	\includegraphics*[width=9.2 cm,clip=true]{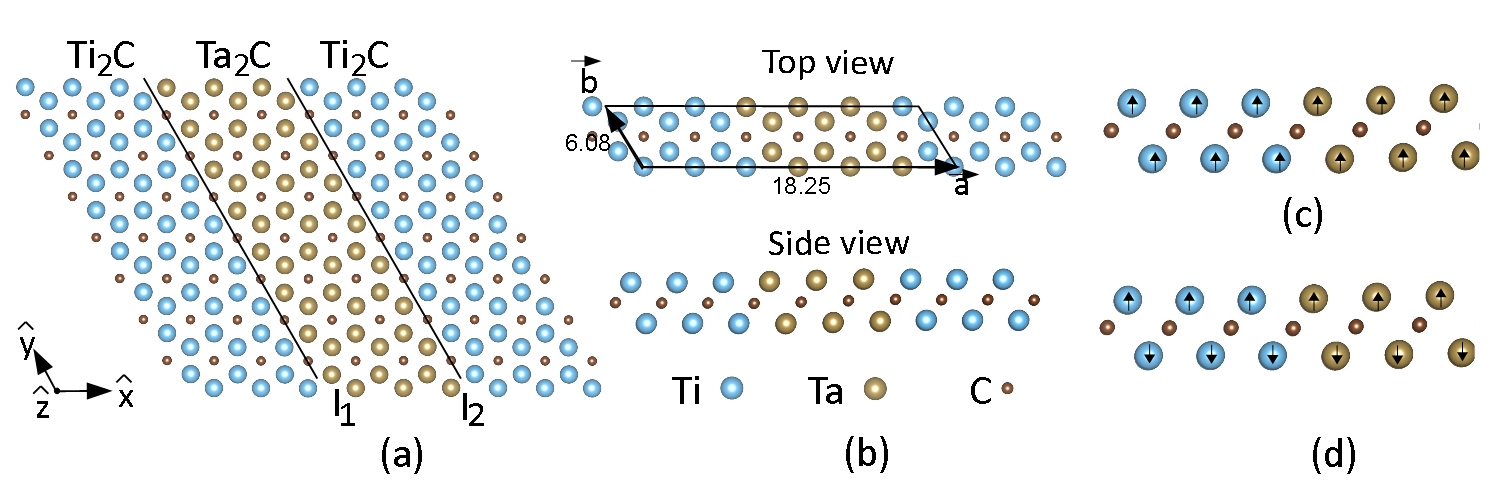}
	\caption{(a) Top and (b) side views of the atomic model of the (Ti$_2$C)$_p$/(Ta$_2$C)$_q$ lateral
		heterostructure with p = 3 and q = 3. 
	Two-dimensional unit cell with lattice parameters and Bravais lattice vectors \textbf{a} and \textbf{b} are also indicated. I$_1$ and I$_2$ are interline (or boundaries, interfaces) between different constituent MXene stripes. (c–d) Side views of the (Ti$_2$C)$_3$/(Ta$_2$C)$_3$ showing the ferromagnetic (FM) and antiferromagnetic (AFM) states, respectively.} 
	\label{models}
\end{figure}

\begin{table*}
	\centering 
	\caption{Optimized values of the lateral heterostructures (Ti$_2$C)$_p$/(Ta$_2$C)$_q$ for $p$ =$q$ (2,3,4,5,10), lattice constants $a$ and $b$ (in {\AA}), and formation energy E$_f$ (in eV).}
		\centering \vspace{2mm} \centering
	\footnotesize\setlength{\tabcolsep}{8pt}
	\begin{tabular}{c| c c c}
		\hline  
		
		~Composite (p/q) ~&~$a~$ &~$b~$  &	E$_f$	\\[0.5ex]
		\hline 
		 			
		(2/2)    & 12.15 &	6.08 & 0.29	  \\[0ex]
		
		(3/3)  & 18.25 &	6.08&	-0.67 \\[0ex]\raisebox{1.5ex}
		
		(4/4) & 24.35 &	6.09 & -0.51  \\[0ex]\raisebox{1.5ex}
		
		(5/5) & 30.44 &	6.085 &	-0.41 \\[0ex]\raisebox{1.5ex}
		
		(10/10)  & 60.94 &	6.09 &	-0.20  \\
	\hline
\end{tabular}
\end{table*}

We first evaluated the stability of all the structures by analyzing the formation energy (E$_f$), following the equation

\begin{equation}
\begin{gathered}
$E$_f$ = ($E_{total}$ – $E_{tot}$(Ti$_2$C) – $E_{tot}$(Ta$_2$C))/N$ 
\end{gathered}
\end{equation}

where $E_{total}$, $E_{tot}$(Ti$_2$C), and $E_{tot}$(Ta$_2$C) represent the total energy of (Ti$_2$C)$_p$/(Ta$_2$C)$_q$ LH, total energies of Ti$_2$C and Ta$_2$C monolayers, respectively, and $N$ is the total number of atoms in the LH. We summarize the optimized values of the formation energy and lattice constants in Table 1. According to our notation, a more negative formation energy indicates a more stable structure. The negative formation energies indicate that the synthesis of all the studied LH (from p=q=3 to 10) is highly possible and most likely to be achieved experimentally than that of those structures with positive formation energies. As seen in Table 1, as the value of $p$=$q$ increases the absolute value of the formation energy decreases. Similar behavior has also been found for other LH, such as the LH based on boron phosphide (BP) and GaN monolayers \cite{Wang21}.

\begin{table*}
\centering 
 \caption {Calculated elastic constants: Elastic constants ($C_{ij}$ in N/m), 
Hill Shear modulus (G in N/m), Hill Bulk modulus (K) (N/m), G/K ratio, and Poisson’s ratios ($\nu$) of Ti$_2$C, Ta$_2$C, and (Ti$_2$C)$_3$/(Ta$_2$C)$_3$ LH.}
 	
 	\centering \vspace{2mm} \centering
 \footnotesize\setlength{\tabcolsep}{4pt}
\begin{tabular}{c| c c  c c c  c  c }
	\hline &&&&&&& \\[-3ex]
	
		~ Structure	& 	$C_{11}$ &	$C_{12}$ &	$C_{66}$ &	
  	$G$  &	$K$ & $G/K$ &	$\nu$ \\
	
	\hline
	Ti$_2$C &	160 &	59 &	50.74&	
 	50.74 &	109.22 &	0.46 &	0.37  \\[0ex]
	
	Ta$_2$C&		235 &	136&	50&	
 	49.55&	185.20&	0.27&	0.57 \\[0ex]
(Ti$_2$C)$_3$/(Ta$_2$C)$_3$&	217	&91&	64&	
	63.31&	154.07&	0.41&	0.40 (Min)/0.43 (Max)   \\[0ex]
	
		\hline
\end{tabular}
\end{table*}

To explore the mechanical stability and the potential performance under different stress conditions, we investigated the mechanical properties of Ti$_2$C, Ta$_2$C, and (Ti$_2$C)$_3$/(Ta$_2$C)$_3$ LH using the stress-strain approach ~\cite{Le-02}. According to the Born stability criteria~\cite{Born-08}, 2D hexagonal crystals should obey the following conditions: 
\begin{equation}
	\label{eq2}
	\ C_{11}>0, \;\ C_{66}>0, \;\
	\ 2*C_{66}=C_{11}-C_{12}, \;\  and \ C_{11}> |C_{12}| 
\end{equation}
The calculated in-plane elastic constants for (Ti$_2$C)$_3$/(Ta$_2$C)$_3$ LH is listed in Table 2. For comparison, we also show the results obtained for the Ti$_2$C and Ta$_2$C MXenes. Our results demonstrate that all the calculated heterostructures satisfy the mechanical stability criteria, and they are thus mechanically stable. The in-plane stiffness (C) can be calculated as 
\begin{equation}
	\label{eq3}
 \begin{split}
C = C_{11} * (1 - ( C_{12}/C_{11})^2)
\end{split}
\end{equation}
and it is a measure of the monolayer’s mechanical strength. Here, the two independent elastic constants, C$_{11}$ and C$_{12}$, can also be used to evaluate the elastic properties of homogeneous and isotropic materials. The in-plane stiffness of (Ti$_2$C)$_3$/(Ta$_2$C)$_3$ LH is
179 N/m, which is higher than the values of both isolated Ti$_2$C (138 N/m) and Ta$_2$C (156 N/m). This result implies that the lateral combination of two monolayers leads to a stronger 2D material. To put this value in context, we note that, according to our results, the mechanical strength of (Ti$_2$C)$_3$/(Ta$_2$C)$_3$ LH is smaller than that of graphene (340.8 N/m)~\cite{Peng-13} but higher than that of MoS$_2$ (120.1 N/m)~\cite{Peng-13}, WS$_2$ (106.4 N/m)~\cite{Zhang21}, Silicene (61.34 N/m)~\cite{John16}, or Germanene (42.05 N/m)~\cite{John16}.

Bulk modulus and shear modulus describe, respectively, the resistance of a material to volume and shape change. To obtain these values, the Voigt (upper bound of elastic properties in terms of uniform strain) ~\cite{Voigt-28}, Reuss (lower bound in terms of uniform stress) ~\cite{Reuss-29}, and Hill (the average between Voigt and Reuss) ~\cite{Hill-52} approximations are mainly used. The ratio of shear to bulk modulus (G/K) roughly reflects the ductility of the material. If the G/K value is less than 0.5, then the material is expected to be ductile ~\cite{Li-09}. From our results we infer, then, that all the calculated MXenes exhibit ductility, a behavior also shown by some 2D materials such as germanene and stanene ~\cite{John16, Lee-08}, in contrast to, for instance, graphene and silicene, which are known to be brittle ~\cite{John16}. Besides the G/K ratio, Poisson’s ratio also gives an idea of the brittle/ductile behavior of materials. The bigger the Poisson’s ratio, the better the material plasticity. For example, while $\nu$ $\sim$ 0.1 indicates more brittleness (such as for pure covalent materials, as is the case of graphene), a typical metallic behavior with $\nu$ = 0.33 exhibits ductility, as the aforementioned example of germanene ~\cite{John16}. As shown in Table 2, the Poisson’s ratio of all calculated MXenes is larger than 0.33, indicating the ductility of these materials.

%

\begin{figure}[h] \centering
	\includegraphics*[width=9cm,clip=true]{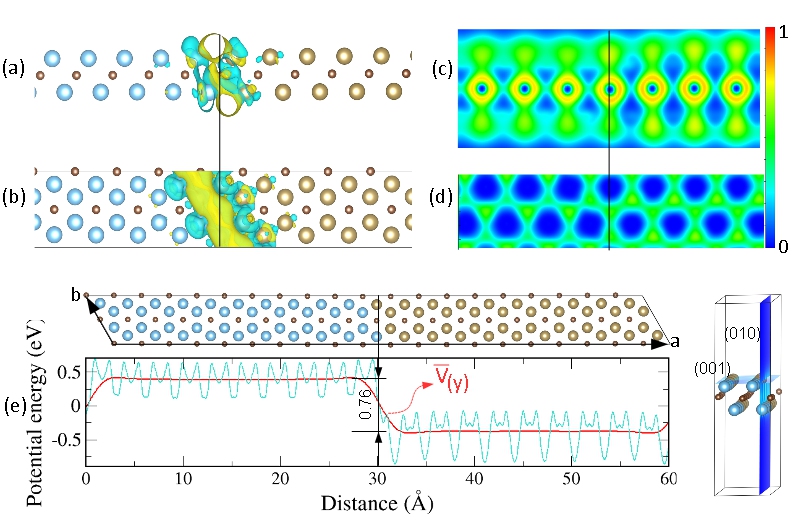}
	\caption{(a) Side and (b) top views of the charge density difference for the (Ti$_2$C)$_{10}$/(Ta$_2$C)$_{10}$LH. The value of the depicted isosurfaces is 0.008 e/${\AA}^3$. The yellow and cyan regions indicate accumulation and depletion of charge, respectively. Electron localization functions of the (Ti$_2$C)$_{10}$/(Ta$_2$C)$_{10}$ LH along the (c) (010) plane and (d) (001) plane. The (010) and (001) planes are shown in the bottom right corner. (e) Planar averaged electrostatic potential energy and its average along the y-direction, $\overline{V(y)}$.} 
	\label{models}
\end{figure}
 
In order to get some insight into the electronic properties, we have analyzed the charge redistribution across the LH. For the smallest LH studied here, i.e. (3,3) and (4,4), the charge density redistributes homogeneously in the Ti$_2$C side and there is no accumulation of charge at the interface (see Fig.S1 in the Supplementary Information). We have thus focused on the largest structure ((Ti$_2$C)$_{10}$/(Ta$_2$C)$_{10}$) to obtain a better understanding of the charge density evolution away from the boundaries. The charge density difference $\Delta\rho(\bf{r})$ is obtained as:
\begin{equation}
	\label{eq2}
	\Delta\rho(\textbf{r})=\rho_{(Ti_2C)_{10}/(Ta_2C)_{10}}(\textbf{r})-
	[\rho_{(Ti_2C)_{10}}(\textbf{r})+\rho_{(Ta_2C)_{10}}(\textbf{r})]
\end{equation}

where $\rho_{(Ti_2C)_{10}/(Ta_2C)_{10}}(\textbf{r})$ represents the total charge density of (Ti$_2$C)$_{10}$/(Ta$_2$C)$_{10}$ LH, and $\rho_{(Ti_2C)_{10}}(\textbf{r})$ and $\rho_{(Ta_2C)_{10}}(\textbf{r})$ are the charge densities of the isolated $(Ti_2C)_{10}$ and $(Ta_2C)_{10}$ monolayers, respectively. The three-dimensional charge density difference of (Ti$_2$C)$_{10}$/(Ta$_2$C)$_{10}$ LH is plotted in Fig. 2. (a-b), where the yellow and cyan regions indicate accumulation and depletion of charge, respectively. As it can be seen, the charge accumulates in the interline region of the heterostructure over an area of approximately 6 {\AA}, while there is almost no charge transfer at either Ti$_2$C or Ta$_2$C far away from the interline because of the weak interaction between the Ti$_2$C and Ta$_2$C monolayers. This interface charge rearrangement is due to the electronegativity difference between the interface atoms. Since the Ti atom is more electronegative (1.54) than the Ta atom (1.50), some electrons flow from the Ti$_2$C monolayer to the Ta$_2$C monolayer when the two materials are brought together, leading to a charge accumulation in the region near the interline. To provide a more quantitative insight and determine the amount of charge transfer between Ti$_2$C and Ta$_2$C we have employed the Bader charge analysis, which gives the result of a charge transfer of 3.7e for the (Ti$_2$C)$_{10}$/(Ta$_2$C)$_{10}$ LH.

%

The electron distribution can be further analyzed by looking at the electron localization functions (ELF) ~\cite{Silvi-94}. Theoretically, an ELF value of 0.5 indicates that the Pauli repulsion has the same value as that in a uniform electron gas of the same density ~\cite{Koh-02}, whereas ELF=1 (0) corresponds to perfect localization (delocalization), where the bonding would have a covalent (metallic) character ~\cite{Shi-17, Liu-19}. From the ELF map in Fig 2 (c-d), we can see that the (Ti$_2$C)$_{10}$/(Ta$_2$C)$_{10}$ LH surface shows a cloud of lone-pair electrons, the density of electrons at the Ta$_2$C side being higher than at the Ti$_2$C side. The substantial concentration of electrons is mostly located around the C atoms, which implies the presence of a covalent bond between the C-C atoms. Similar behavior was also found for Ti$_4$C$_3$ and Ti$_4$N$_3$ ~\cite{Zhang18}.

Fig. 2(e) shows as well that both the planar average electrostatic potential and its average value $\overline {V(y)}$ changes at the boundaries of LH, which can be explained in terms of the difference in the electrostatic potentials of the isolated monolayers. The Ti$_2$C monolayer has a higher electrostatic potential than the Ta$_2$C, with a potential difference calculated to be 0.76 eV. This potential energy difference will induce a built-in electric field, which results in a band bending near the boundaries of the (Ti$_2$C)$_{p}$/(Ta$_2$C)$_{q}$ LH. The workfunction (WF, or $\phi$) is an important physical parameter of materials, and it is defined as the energy required to remove an electron from the Fermi level to vacuum. Low WFs material can thus improve the power efficiency of a device. We have calculated the WF of Ti$_2$C, Ta$_2$C, and (Ti$_2$C)$_{10}$/(Ta$_2$C)$_{10}$ LH as 4.67 eV, 4.97 eV, and 4.80 eV, respectively. The values found for the WF of Ti$_2$C and Ta$_2$C are in good agreement with other theoretical calculations (Ti$_2$C (4.53 eV)~\cite{Thomas21} and Ta$_2$C (4.77 eV)~\cite{Liu16}). Hence, the (Ti$_2$C)$_{10}$/(Ta$_2$C)$_{10}$ LH presents a lower WF than that of graphene/black phosphorene bilayer (5.38 eV)~\cite{Cai-15} and Ti$_2$CO$_2$ (5.90 eV) ~\cite{Li-20}, making it potentially more promising to improve the performance of nanoelectronic devices. The WF can also explain the origin of the charge accumulation at the interface. It is known that when two metals are brought into contact no energy barrier will arise at the interface, even if their workfunctions are different ~\cite{Giu-17}. If the electrons flow from a small (big) workfunction metal to another, there will be a positive (negative) charge space in the second metal near the interface with the first one ~\cite{Giu-17}. In our case, the charge transfer occurs from the Ti$_2$C (small workfunction side) to the Ta$_2$C monolayer, and hence electrons accumulate at the interface region of the (Ti$_2$C)$_{10}$/(Ta$_2$C)$_{10}$ LH on the Ta$_2$C side, while there is a charge depletion on the Ti$_2$C side of the LH, which is consistent with Fig. 2. (a-b).

Finally, we have explored the magnetic properties of these LH. Ongoing research in spintronics is focused on developing new materials and devices that can better harness the spin of electrons and exploring new physical phenomena that arise from spin transport and manipulation. In particular, finding novel 2D materials with high Curie temperatures is essential to bring spintronics from the laboratory to commercial applications. We start our investigation by performing a spin-polarized geometry optimization of the isolated 2D MXenes, Ti$_2$C and Ta$_2$C, and the LH. The spin-polarized optimized lattice parameters of Ti$_2$C and Ta$_2$C are 3.086 {\AA} and, 3.081 {\AA}, respectively. Ti$_2$C presents an interesting magnetic ground state, with a non-zero magnetic moment in the ferromagnetic (FM) but whose ground state is antiferromagnetic (AFM) (magnetic moment zero), while Ta$_2$C is non-magnetic. Their respective magnetic moments are 1.90  $\mu$B for Ti$_2$C and 0.00 $\mu$B for Ta$_2$C. These results are in good agreement with previous studies, which found 3.083 {\AA} and 1.91 $\mu$B for Ti$_2$C~\cite{Gao-16, Rom23} and 3.083 {\AA} and 0.00 $\mu$B for Ta$_2$C ~\cite{Zhao14}. The optimized lattice constants for the (Ti$_2$C)$_p$/(Ta$_2$C)$_q$ LH were obtained by relaxing both the lattices and the atomic positions for the magnetic ground state, and are given in Table 3. 

The next step is to determine the magnetic ground state of each LH (Ti$_2$C)$_p$/(Ta$_2$C)$_q$ for ($p$ =$q$ = 3, 4, 5, 10), considering the FM, AFM, and nonmagnetic (NM) states, as showcased in Fig. 1. Table 3 displays the values obtained for the total energy for the three possible configurations for all the studied LH. The FM state has the lowest total energy compared to the NM and the AFM states for all structures, and it is thus the magnetic ground state. The band structure of the 
(Ti$_2$C)$_3$/(Ta$_2$C)$_3$ (Fig.S2 in Supplementary Information) 
shows dispersive metallic bands crossing the Fermi level (E$_F$). Such bands are actually the signature of a ferromagnetic metal behavior, according to the Stoner theory ~\cite{Capellmann79}. The analysis of the energy bands and the density of states (DOS) of the (Ti$_2$C)$_3$/(Ta$_2$C)$_3$ that are depicted in Fig.3 shows that the predominant contribution to the dispersing bands that cross the E$_F$ comes the Ti d-orbitals. This means that the Ti d electronic states around the E$_F$ are indeed itinerant, thus favoring ferromagnetism according to the Stoner theory. This behavior is also found for other 2D MXenes such as Cr$_2$C ~\cite{Si15} and Ti$_2$C ~\cite{Akgenc20}.

The spin-resolved (DOS) of the FM state for the (Ti$_2$C)$_3$/(Ta$_2$C)$_3$ LH provides further insight into the magnetic nature of the LH. As can be seen in Fig. 3, the heterostructure remains metallic, the metallicity mainly originating from the d orbital of the Ti and Ta atom in (Ti$_2$C)$_3$/(Ta$_2$C)$_3$ LH, as it happens for the isolated structures.

\begin{figure}[h] \centering
	\includegraphics*[width=9 cm,clip=true]{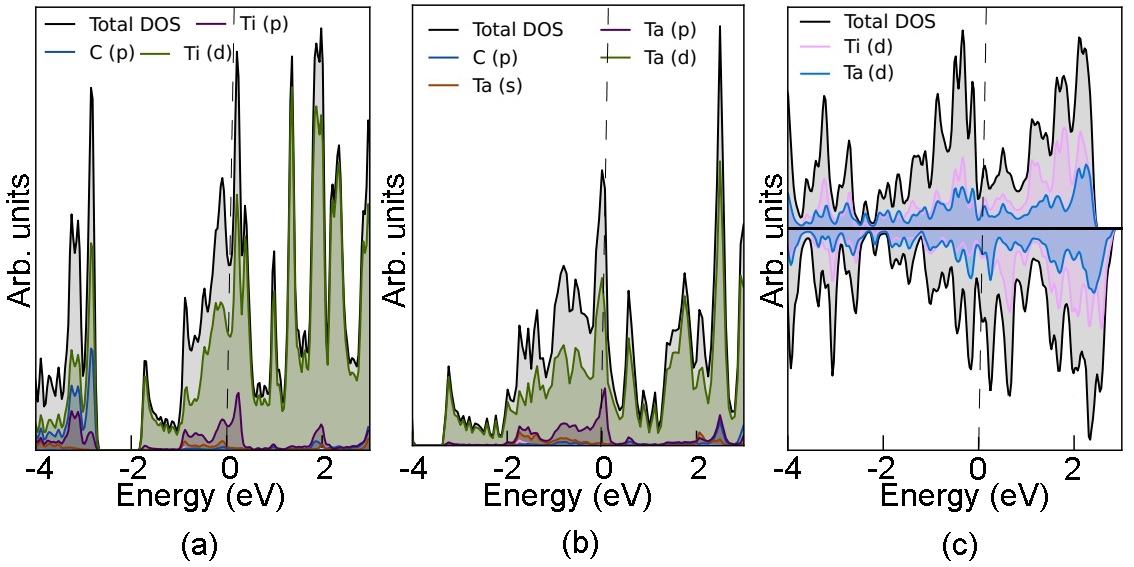}
	\caption{Projected Density of States (PDOS) of the (a) Ti$_2$C, (b)Ta$_2$C, (c) Spin-resolved density of states of the (Ti$_2$C)$_3$/(Ta$_2$C)$_3$ LH for the p and d orbitals of the Ti and Ta atoms. The Fermi level is set at 0 eV.} 
	\label{models}
\end{figure}

\begin{table*}
	\centering 
	\caption{The optimized lattice constants a$_0$ (in \AA) with the FM state, the total ($M_{tot}$) magnetic moments (in $\mu$B) at equilibrium lattices, the total energy differences of FM with NM (EFM–ENM) and FM with AFM (EFM–EAFM) states (in eV), and the Curie temperatures (K) and Magnetic Anisotropy Energy (MAE) ($\mu$eV per (Ti+Ta) atom) for the high symmetry directions for (Ti$_2$C)$_p$/(Ta$_2$C)$_q$ for $p$ = $q$ (3,4,5,10)}
		\centering \vspace{2mm} \centering
	\footnotesize\setlength{\tabcolsep}{8pt}
	\begin{tabular}{c| c c c c c c c c c c}
		\hline  
		
		~Composite (p/q) ~&~$a~$ &~$M_{tot}$  &	E$_{FM-NM}$	&	E$_{FM-AFM}$ & T$_C$ & (100) & (010) & (110) & (001) & (111)  \\[0.5ex]
		\hline 
		
		(3/3)  & 18.34 &	10.22 & -0.39 &  -0.09 & 696 & 
  19&	6	&0.00&	58&	32 \\[0ex]\raisebox{1.5ex}
		
		(4/4) & 24.47 &	13.67 & -0.54 &  -0.14 & 1082&  20	&5&	0.00	&104&	38 \\[0ex]\raisebox{1.5ex}
		
		(5/5) & 30.58 &	17.85 &	-0.69 & -0.11 & 850&  
  21&	6	&0.00&	106&	38\\[0ex]\raisebox{1.5ex}
		
		(10/10)  & 61.21 &	36.18 &	-1.40 & -0.09 & 696&  9&	3&	0.00&	25&	9  \\
	\hline
\end{tabular}
\end{table*}

As we discussed, the Curie or Néel temperature (T$_C$ or T$_N$) is an essential parameter to assess realistically the possibility of using a material in future nanoelectronic devices. A T$_C$ beyond room temperature is a highly demanded characteristic, and there has been much research \cite{You19, You20} trying to find materials with high enough T$_C$. We have then calculated the T$_C$ for all the (Ti$_2$C)$_p$/(Ta$_2$C)$_q$ ($p$ =$q$ = 3, 4, 5, 10) LH by means of the mean-field theory approximation (MFA), which estimates it as k$_B$T$_C$ = (2/3) $\Delta({E})$ ~\cite{Jin-09}. Here, k$_B$ is the Boltzmann constant, and $\Delta({E})$ is the energy difference between FM and the AFM states. The results are summarized in Table 3. The T$_C$ of (Ti$_2$C)$_p$/(Ta$_2$C)$_q$ for $p$ =$q$ (3,4,5,10) LH are as high as 696 K, 1082 K, 850 K, and 696 K, respectively, thus well above room temperature (RT). Even though considering the possible overestimation of MFA, it is widely accepted that it provides a good qualitative variation trend of the T$_C$ ~\cite{Tandon08, Zhang-21, Li21, Yang-2016}, and the estimated values in our case clearly indicate a highly likely high Curie temperature for (Ti$_2$C)$_p$/(Ta$_2$C)$_q$. For comparison, our calculation of the Néel temperature of Ti$_2$C s 247 K, in agreement with previous studies \cite {Rom23}. While 2D ferromagnetism has been experimentally demonstrated for the Cr$_2$Ge$_2$Te$_6$ bilayer and the CrI$_3$ monolayer ~\cite{Gong17, Huang17}, the T$_C$s is that 30 and 45 K, respectively, far below room temperature. Our results, however, corroborate the suitability of other 2D materials for spintronic devices operating at room temperature: Yang {\it et al.} found that both Cr$_2$VC$_2$F$_2$ and Cr$_2$VC$_2$(OH)$_2$ present Curie temperature values of 695.65 K and 618.36 K, respectively ~\cite{Yang-2016}. Such high Curie temperatures (1877 to 566 K) were also found for other 2D MXenes such as Mn$_2$NT$_x$ with different surface terminations (T = O, OH, and F) ~\cite{Kumar-17}, showing the great potential of MXene-based devices for spintronics.    
 
Lastly, we will discuss the magnetic anisotropy energy (MAE) to further explore the FM character of (Ti$_2$C)$_p$/(Ta$_2$C)$_q$ LH. A large MAE implies, in principle, a better resistance of the magnetic ordering against heat fluctuations at the operating temperatures of commercial devices, and many efforts are nowadays devoted to improving the values of MAE in potentially magnetic 2D materials.
The MAE in FM 2D materials emerges mostly from the spin–orbit coupling (SOC) interactions, and as such is examined by considering
the spin-orbit coupling (SOC) along five magnetization directions in the $ab$-plane, i.e., the (100), (010), and (110) directions, and out of the $ab$-plane, i.e. the (001) and (111) directions. The difference in energy between the system with a given spin orientation and the energy of the most stable spin orientation (easy axis) defines the MAE. As seen in Table 3., (Ti$_2$C)$_p$/(Ta$_2$C)$_q$ for ($p$ =$q$ = 3, 4, 5, 10) LH exhibit an in-plane easy axis, which is the (110).  The values of the MAEs for the other SOC orientations are also shown in Table 3 for comparison. The large MAE mainly originates from the strong Ti–Ti and Ta-Ta coupling in the $ab$-plane of (Ti$_2$C)$_p$/(Ta$_2$C)$_q$ ($p$ =$q$ = 3, 4, 5,10) LH. Our results are similar to previous results found for MnP \cite{Wang19}, VS$_2$ \cite{Yue17}, and Mn$_2$C \cite{Hu16}, but larger than those found for Fe (1.4 $\mu$eV per Fe atom) and Ni (2.7 $\mu$eV per Ni atom) bulks \cite{Daalderop90}. The high T$_C$ (and also high magnetic moments), and high MAE of (Ti$_2$C)$_p$/(Ta$_2$C)$_q$ make them promising 2D magnetic materials for spintronic, photocatalysis, and data storage applications at room temperature.

\section{Summary}
Using DFT based-simulations, we have illustrated the prospective mechanic, electronic, and
magnetic properties of the (Ti$_2$C)$_p$/(Ta$_2$C)$_q$ LH. The formation energies of the (Ti$_2$C)$_p$/(Ta$_2$C)$_q$ LH confirm the energetic stability of this 2D material. The investigation of their mechanical properties shows that the lateral combination of two monolayers results in a stronger 2D material than the isolated monolayers. For wide enough LH, most electrons tend to accumulate at the interface, with a spatial charge extension of $\sim$~6 {\AA}. All the studied LH show high Curie temperatures, well beyond room temperature, and high magnetic anisotropy energies. Our findings show the need to reinforce further exploration of MXene-based 2D heterostructures to exploit their elevated capabilities for magnetic storage and spintronic devices.

\section*{Author Contributions}
Sibel Özcan: Conceptualization (lead); formal analysis (lead); writing – original draft (lead); writing – review and editing (equal). Blanca Biel: Conceptualization (supporting); writing – review and editing (equal); funding acquisition (lead).

\section*{Conflicts of interest}
There are no conflicts to declare.

\section*{Acknowledgments}

B.B. kindly acknowledges financial support from Programa 
Operativo FEDER B.FQM.272.UGR20 and AEI under project 
PID2021-125604NB-I00. S.Ö. and B.B. also acknowledge 
financial support by the Junta de Andalucía under the 
Programa Operativo FEDER P18-FR-4834The Albaicín 
supercomputer of the University of Granada and 
TUBITAK ULAKBIM, High Performance and Grid 
Computing Center (TRUBA resources) are also acknowledged 
for providing computational time and facilities.

\end{document}